\documentclass[12pt]{article}
\usepackage{graphicx}
\usepackage{subfig}
\usepackage{epstopdf}
\usepackage{verbatim}
\usepackage{amsmath}
\usepackage{amssymb}
\usepackage{hyperref}
\usepackage{amsfonts}
\newcommand{\ud}{\mathrm{d}}
\newcommand{\ue}{\mathrm{e}}

\newcommand{\D}[2]{\frac{\ud{#1}}{\ud {#2}}}

\newcommand{\pD}[2]{\frac{\partial{#1}}{\partial{#2}}}

\newcommand{\abs}[1]{\left\vert{#1}\right\vert}
\newcommand{\ib}[2]{\mathbf{i}_{\mathbf{#1}}^{#2}}
\newcommand{\jb}[2]{\mathbf{j}_{\mathbf{#1}}^{#2}}
\newcommand{\eb}[2]{\mathbf{e}_{\mathbf{#1}}^{#2}}
\newcommand{\norm}[1]{\left\|{#1}\right\|}
\renewcommand{\Re}[1]{\mathbb{R}\text{e}\de{#1}}

\newcommand{\Hy}[1]{\mathbb{H}\text{y}\de{#1}}
\newcommand{\bo}[1]{\boldsymbol{#1}}
\newcommand{\wh}[1]{\widehat{#1}}
\newcommand{\ke}[1]{\vert{#1}\rangle}

\newcommand{\cro}[1]{\left[ {#1} \right]}
\newcommand{\acco}[1]{\left\lbrace {#1} \right\rbrace }
\newcommand{\paren}[1]{\left( {#1} \right)}
\newcommand{\p}[1]{\left( {#1} \right)}
\newcommand{\de}[1]{\!\paren{#1}}
\newcommand{\ol}[1]{\overline{#1}}
\newcommand{\ie}{\textit{i.e.} }

\newcommand{\al}{\alpha}
\newcommand{\be}{\beta}

\newcommand{\del}{\delta}

\newcommand{\ep}{\epsilon}

\newcommand{\ze}{\zeta}

\def\R{\mathbb{R}}
\def\C{\mathbb{C}}
\def\T{\mathbb{T}}
\newcommand{\Title}[1]{\large\textbf{#1}\normalsize}
\newcommand{\Author}[1]{\textrm{#1}}
\newcommand{\Institution}[1]{\textit{#1}}
\newcommand{\Email}[1]{\texttt{#1}}
\begin{document}
\begin{center}
\Title{\scalebox{1.15}{The bicomplex quantum Coulomb}}\vspace{0.25cm}\\
\Title{\scalebox{1.15}{potential problem}}\vspace{0.75cm}\\
\Author{J\'er\'emie Mathieu,$^1$ Louis Marchildon$^2$ and Dominic Rochon$^3$}\\[0.3cm]
\Institution{$^1$D\'epartement de physique,\\ Universit\'e
de Montr\'eal, Montr\'eal, Qc, Canada, H3C 3J7}\\
\Email{\small jeremie.mathieu@umontreal.ca}\\[0.3cm]
\Institution{$^2$D\'epartement de chimie, biochimie et
physique,\\ Universit\'e du Qu\'ebec, Trois-Rivi\`eres, Qc, Canada, G9A 5H7}\\
\Email{\small louis.marchildon@uqtr.ca}\\[0.3cm]
\Institution{$^3$D\'epartement de math\'ematiques et d'informatique,\\
 Universit\'e du Qu\'ebec, Trois-Rivi\`eres, Qc, Canada, G9A 5H7}\\
\Email{\small dominic.rochon@uqtr.ca}\\[0.5cm]
\end{center}
\begin{abstract}
\begin{sloppypar}
Generalizations of the complex number system underlying the mathematical formulation of quantum mechanics have been known for some time, but the use of the commutative ring of bicomplex numbers for that purpose is relatively new. This paper provides an analytical solution of the quantum
Coulomb potential problem formulated in terms of bicomplex numbers. We define the problem by introducing a bicomplex hamiltonian operator and extending the canonical commutation relations to the form $\cro{X_i,P_k}=\ib{1}{}\hbar\xi\del_{ik}$, where $\xi$ is a bicomplex number. Following Pauli's algebraic method, we find the eigenvalues of the bicomplex hamiltonian. These eigenvalues are also obtained, along with appropriate eigenfunctions, by solving the extension of
Schr\"odinger's time-independent differential equation. Examples of solutions are displayed. There is an orthonormal system of solutions that belongs to a bicomplex Hilbert space.
\end{sloppypar}
\end{abstract}

\section{Introduction}\label{introduction}

It is generally believed that the best justification of a physical theory, no matter what the nature of its mathematical formalism, rests on the agreement of its predictions with experiment and the internal consistency or elegance of the theory
itself~\cite{Reichenbach1944philosophic}. Views on the mathematical formalism range from pure instrumentalism to the idea that the formalism itself is
real~\cite{Penrose2005road}. 

\begin{sloppypar}
The mathematical formalism of quantum mechanics has been studied
thoroughly~\cite{neumann1955mathematical,jauch,prugovecki}.  Not everyone agrees on the set of postulates necessary to build the foundations of a coherent quantum
mechanics~\cite{hardy,volovich2002seven}. The Hilbert space structure of the set of quantum states, however, seems to be
uncontroversial.
\end{sloppypar}

Hilbert spaces used in quantum mechanics are defined over the nonordered field of complex 
numbers~$\C$. Complex numbers make up a division algebra richer than their real subset, and are deeply connected with superposition of
quantum-mechanical amplitudes. Unitary representations of Lie groups, fundamental tools in the quantum theory of
symmetry, require complex numbers in an essential
way~\cite{weyl,wigner}. 

Quantum mechanics postulates that the only possible results of the measurement of a dynamical variable are the eigenvalues of the corresponding
self-adjoint operator acting in the state space. As eigenvalues of
self-adjoint operators are real, this is a form of correspondence between the quantum and classical world descriptions. 

If complex numbers are so appropriate to describe the quantum world, one can ask whether generalizations of that number system might do equally well or even better. The noncommutative field of quaternions has already been investigated from that point of 
view~\cite{adler}. More recently, attention has turned towards the commutative ring of bicomplex
numbers~\cite{Price1991introduction,rochon2004algebraic}. This paper is part of a program of extending the quantum mathematical formalism to that algebraic structure, which is neither a division algebra nor an 
absolute-valued algebra over the real
numbers~\cite{rochon2004bicomplex,rochon2006bicomplex,lavoie2010bicomplex,lavoie2011finite,lavoie2010infinite,lavoie2012uncertainty}.

\begin{sloppypar}
In section~\ref{binumandfcts}, we review some of the algebraic properties of bicomplex numbers and construct an
infinite-dimensional bicomplex Hilbert space made up of
square-integrable bicomplex functions.
Section~\ref{coupotprob} defines the bicomplex generalization of the
quantum-mechanical Coulomb potential problem. Eigenvalues of the corresponding hamiltonian are obtained through Pauli's algebraic method.
Section~\ref{eigenfctsofH} is devoted to obtaining eigenfunctions of the bicomplex Coulomb potential hamiltonian in the coordinate basis. To our knowledge, this is the first time that this has been done with an algebra larger than $\C$.  Graphical representations of some functions of interest are shown in
section~\ref{graphrep}. In section~\ref{veriandconclu} we check the consistency of some of the assumptions made and show that the Coulomb eigenfunctions live in a bicomplex Hilbert space.  We conclude in section~\ref{conclu}.
\end{sloppypar}

%\newpage
\section{Bicomplex numbers and functions}\label{binumandfcts}
\begin{sloppypar}
In this section we briefly summarize relevant algebraic properties of bicomplex numbers.  More information and proofs can be found
in~\cite{Price1991introduction,rochon2004algebraic}. We define the concept of a
square-integrable bicomplex function.  Such functions are then used to construct an
infinite-dimensional bicomplex Hilbert space which, it turns out, will be an appropriate arena for bicomplex Coulomb potential eigenfunctions.
\end{sloppypar}

\subsection{Algebraic structure}\label{algstruc}

One way to define a bicomplex number $\al$ is by writing
\begin{align}
\al:=\al_{\wh{1}}\eb{1}{}+\al_{\wh{2}}\eb{2}{},\label{fbih_eq:defbi}
\end{align}
where $\al_{\wh{1}}$ and $\al_{\wh{2}}$ belong to $\C\de{\ib{1}{}}$, the field of complex numbers. The caret
notation~\cite{lavoie2010bicomplex} is used to label complex components of bicomplex numbers, thereby avoiding confusion with other kinds of indices. The imaginary bicomplex units $\eb{1}{}$ and $\eb{2}{}$ satisfy the remarkable properties
\begin{align}
\eb{1}{2}=\eb{1}{}, \quad \eb{2}{2}=\eb{2}{},
\quad \eb{1}{}+\eb{2}{}=1, \quad \eb{1}{}\eb{2}{}=0=\eb{2}{}\eb{1}{}.
\label{fbih_eq:defbasebi}
\end{align}
We call $\acco{\eb{1}{},\eb{2}{}}$ the idempotent basis of bicomplex numbers. With the addition and multiplication defined in the obvious way, the set of bicomplex numbers $\T$ forms a commutative ring with unity. Properties~\eqref{fbih_eq:defbasebi} greatly simplify bicomplex algebraic calculations and make definition~\eqref{fbih_eq:defbi}, among other equivalent choices, a very useful one for our purposes. 

If $\al_{\wh{1}}=0$, then~\eqref{fbih_eq:defbasebi} implies that $\alpha \eb{1}{} = 0$.
In fact any bicomplex number $\al$ for which either $\al_{\wh{1}}=0$ or $\al_{\wh{2}}=0$ is a zero divisor.
The set of all zero divisors is called the null cone (not to be confused with the light cone of special relativity) and is denoted by~$\mathcal{NC}$.  Idempotents $\eb{1}{}$ and $\eb{2}{}$ project bicomplex numbers onto complementary minimum ideals.

Define $\jb{}{}$, the imaginary hyperbolic unit, as $\jb{}{}:=\eb{1}{}-\eb{2}{}$.
Then $\jb{}{2}=1$, $\eb{1}{}=\p{1+\jb{}{}}/2$ and $\eb{2}{}=\p{1-\jb{}{}}/2$. Substituting this
in~\eqref{fbih_eq:defbi}, we get the hyperbolic representation of $\al$ as
\begin{align}
\al =\acco{\frac{\al_{\wh{1}}+\al_{\wh{2}}}{2}}+\acco{\frac{\al_{\wh{1}}-\al_{\wh{2}}}{2}}\jb{}{}=:x_{\al}+y_{\al}\,\jb{}{}, \label{eqalpha}
\end{align}
which in turn means that $\al_{\wh{1}}=x_{\al}+y_{\al}$ and $\al_{\wh{2}}=x_{\al}-y_{\al}$. If $\al_{\wh{1}}$ and $\al_{\wh{2}}$ are both
in~$\R$, we call $\alpha$ a hyperbolic number. The set~$\mathbb{D}$ of all hyperbolic numbers is obviously a subset
of~$\T$.  Note that $(-\ib{1}{} \jb{}{})^2 = -1$, so that $-\ib{1}{} \jb{}{}$ has the properties of an imaginary unit.  It is usually called $\ib{2}{}$.

There are several ways to define conjugation in~$\T$, but the one most useful for our purposes is the following.  We define $\al^\dagger$ as  $\ol{\al_{\wh{1}}}\,\eb{1}{}+\ol{\al_{\wh{2}}}\,\eb{2}{}$, where the upper bar denotes the usual complex conjugation. Clearly,
\begin{align}
\al^\dagger\al = \abs{\al_{\wh{1}}}^2\eb{1}{}+\abs{\al_{\wh{2}}}^2\eb{2}{},
\end{align}
where $\abs{\ \ }$ is the standard real norm of complex algebra.

The real norm of a bicomplex number $\al$ is defined as
\begin{align}
\abs{\al}:=\frac{1}{\sqrt{2}}\sqrt{\abs{\al_{\wh{1}}}^2+\abs{\al_{\wh{2}}}^2}.\label{fbih_eq:defmodulusbi}
\end{align}
If $\al$ is hyperbolic, then 
\begin{align*}
\abs{\al}=\sqrt{x_{\al}^2+y_{\al}^2}:=\sqrt{\Re{\al}^2+\Hy{\al}^2}.
\end{align*}
One can show~\cite{Price1991introduction} that for all $\al,\be \in\T$ and $z\in\C\de{\ib{1}{}}$,
%
%% One column
%\begin{comment}
\begin{align*}
\abs{\al}\geq0, \quad \abs{z\al}=\abs{z}\abs{\al}, \quad \abs{\al+\be}
\leq\abs{\al}+\abs{\be} \quad \mbox{and} \quad \abs{\al\be}\leq\sqrt{2}\abs{\al}\abs{\be}.
\end{align*}
%\end{comment}
%% two columns
\begin{comment}
\begin{align*}
&\abs{\al}\geq0, \quad \abs{z\al}
=\abs{z}\abs{\al}, \quad \abs{\al+\be}
\leq\abs{\al}+\abs{\be} \\
& \quad \mbox{and} \quad \abs{\al\be}\leq\sqrt{2}\abs{\al}\abs{\be}.
\end{align*}
\end{comment}
%
Since $\T$ possesses zero divisors and since the norm of a product of bicomplex numbers is not in general equal to the product of their respective norms, the algebraic structure $\p{\T,+,\cdot,\abs{\ \ }}$ is neither a division algebra nor an
absolute-valued algebra over the real numbers.

\subsection{Normed function space}\label{normspace}

Defining a bicomplex function $f$ of $q$ bicomplex variables as a $q$-tuple infinite 
positive-integer convergent power series with bicomplex coefficients, one can show that
\begin{align}
f\de{\boldsymbol{\mu}}=f_{\wh{1}}\de{\boldsymbol{\mu}_{\wh{1}}}\eb{1}{}+f_{\wh{2}}\de{\boldsymbol{\mu}_{\wh{2}}}\eb{2}{}.\label{fbih_eq:deffctbidenvarbi}
\end{align}
The notation $f\de{\boldsymbol{\mu}}$ means that $f$ depends on $q$ bicomplex variables $\mu_i$, and each $f_{\wh{s}}$ ($s = 1, 2$) is a $\C\de{\ib{1}{}}$ complex function of the $q$ complex variables $\mu_{i\wh{s}}$. 

We say that $f$ in~\eqref{fbih_eq:deffctbidenvarbi} belongs to the null cone if either $f_{\wh{1}}$ or $f_{\wh{2}}$ is zero. We call $f$ a hyperbolic function if $f_{\wh{1}}$ and $f_{\wh{2}}$ are both real. 

As a particular case, if all $\mu_i$ are real we simply have
\begin{align}
f\de{\boldsymbol{\mu}}=f_{\wh{1}} (\boldsymbol{\mu}) \;\eb{1}{}+f_{\wh{2}} (\boldsymbol{\mu}) \;\eb{2}{}.\label{fbih_eq:deffctbidenvarre}
\end{align}
We say that $f$ is a bicomplex 
square-integrable function if and only if the $f_{\wh{s}}$ are both 
square-integrable functions, that is,
\begin{align}
\int \abs{f_{\wh{s}} \de{\boldsymbol{\mu}}}^2\ud\boldsymbol{\mu}<\infty \label{fbih_eq:intquadfctbi}
\end{align}
for $s=1$ and $2$.  Here $\ud\boldsymbol{\mu}$ is the Lebesgue mesure on~$\R^q$~\cite{appel2007math}. We denote by~$\mathcal{F}_q$ the set of bicomplex
square-integrable functions of $q$ real variables. It can be shown that with standard addition and multiplication, $\mathcal{F}_q$ makes up a $\T$-module. This module is explicitly denoted as $\p{\mathcal{F}_q,\T,+,\cdot}$ and it obviously has infinite dimension. 

\begin{sloppypar}
For any $f,g\in\mathcal{F}_q$, the following binary mapping takes two bicomplex 
square-integrable functions and transforms them into a unique bicomplex number:
\end{sloppypar}
\begin{align}
\p{f,g}&:=\int  f^\dagger\de{\boldsymbol{\mu}}g\de{\boldsymbol{\mu}}\ud\boldsymbol{\mu}
=\sum_s\eb{s}{}\int\ol{f_{\wh{s}}\de{\boldsymbol{\mu}}}g_{\wh{s}}\de{\boldsymbol{\mu}}\ud\boldsymbol{\mu}.\label{fbih_eq:defprodscafctbi}
\end{align}
If we identify functions that differ only on a set of measure zero, the binary
mapping~\eqref{fbih_eq:defprodscafctbi} satisfies all the properties of a scalar product.  Explicitly,
\begin{enumerate}
\item $(f,g+h) = (f,g) + (f,h)$;
\item $(f,\alpha g) = \alpha (f,g)$; 
\item $(f,g) = (g,f)^{\dagger}$;
\item $(f,f) = 0$ if and only if $f=0$.
\end{enumerate}
The functions $f$ and $g$ are orthogonal if their scalar product vanish. We say that $f$ is normalized if $(f,f) = 1$. It follows from the third property that $(f,f)$ is always a hyperbolic number.

With \eqref{fbih_eq:defprodscafctbi}, one can define an induced $\T$-norm on $\mathcal{F}_q$ as
\begin{align}
\norm{f}:=\frac{1}{\sqrt{2}}\sqrt{\p{f,f}_{\wh{1}}+\p{f,f}_{\wh{2}}}
=\frac{1}{\sqrt{2}}\sqrt{\sum_s\int\abs{f_{\wh{s}}\de{\boldsymbol{\mu}}}^2\ud\boldsymbol{\mu}}.\label{fbih_eq:definducednorm}
\end{align}
Making use of results proved in~\cite{lavoie2010infinite}, it is not difficult to show that the structure
$\p{\mathcal{F}_q,\T,+,\cdot,\p{\ ,\ },\norm{\ \ }}$ is a bicomplex Hilbert space.

%\newpage
\section{The Coulomb potential problem}\label{coupotprob}

In standard quantum mechanics, the hamiltonian associated with the Cou\-lomb potential is given by
\begin{align}
H=\frac{1}{2\mu}P^2-\frac{Ze^2}{R},\label{fbih_eq:defHfromXiPk}
\end{align}
where $\mu$, $e^2$ and $Z$ are positive real numbers and
\begin{align}
P^2:=P_1^2+P_2^2+P_3^2 , \qquad R:=\sqrt{X_1^2+X_2^2+X_3^2}.\label{fbih_eq:defHfromXiPk1}
\end{align}
The six operators $X_i$ and $P_k$ are self-adjoint and satisfy the commutation relations $\cro{X_i,P_k}=\ib{1}{}\hbar\del_{ik}$, where on the 
right-hand side the identity operator is implicit.

The quantum-mechanical Coulomb problem consists in finding the eigenvalues and eigenvectors of $H$, that is, it consists in solving the equation $H\ke{\psi_E}=E\ke{\psi_E}$. Its most important application is the determination of the energy levels and state vectors of a hydrogen atom or
hydrogen-like ion in its center-of-mass
frame. The Coulomb problem is one of the few analytically solvable problems of quantum mechanics.  The solution can be obtained both by an algebraic method that goes back to 
Pauli~\cite{pauli,englefield1972group,greiner1994quantum}, and by a differential equation method that goes back to Schr\"{o}dinger~\cite{schrodinger,tannoudji1977quantum,marchildon2002quantum}.

\subsection{Statement of the bicomplex problem}\label{defsandaxioms}

The quantum Coulomb problem will now be formulated in terms of bicomplex numbers.
The crucial step consists in extending the canonical commutation relations the way it was done for the harmonic
oscillator~\cite{lavoie2010bicomplex}. We proceed by making a set of assumptions, from which we will derive a number of properties satisfied by eigenvalues and eigenkets of the hamiltonian. The consistency of the assumptions will eventually be checked through the explicit solutions obtained. So here are our assumptions: 
\begin{enumerate}
\item[a)] Seven bicomplex linear operators $X_i$, $P_k$ and $H$, related
by~\eqref{fbih_eq:defHfromXiPk} and \eqref{fbih_eq:defHfromXiPk1}, act in a $\T$-module $\mathcal{M}$. Elements of $\mathcal{M}$ are called kets and are generically denoted as $\ke{\psi}$.  Operators and kets can be decomposed in the idempotent basis as
\begin{align}
X_i &= X_{i\bo{1}}\eb{1}{}+X_{i\bo{2}}\eb{2}{},\\
\ke{\psi} &= \ke{\psi}_{\bo{1}} \eb{1}{} +\ke{\psi}_{\bo{2}}\eb{2}{}.
\end{align}
Here $X_{i\bo{s}}:=\eb{s}{}X_i$ and $\ke{\psi}_{\bo{s}}:=\eb{s}{}\ke{\psi}$. We say that $X_i$ belongs to the null cone if either $X_{i\bo{1}}$ or $X_{i\bo{2}}=0$. The same applies to $P_k$ and $H$. Similarly, $\ke{\psi}\in\mathcal{NC}$ if $\ke{\psi}_{\bo{1}}$ or $\ke{\psi}_{\bo{2}}=\ke{0}$.  The bold index notation is such that any quantity affected by a bold subscript $\bo{s}$ is unchanged when multiplied by $\eb{s}{}$.
\item[b)] The operators $X_i$ and $P_k$ are self-adjoint with respect to a bicomplex scalar product to be specified explicitly. The scalar product has to satisfy properties analogous to the ones enumerated after eq.~\eqref{fbih_eq:defprodscafctbi}.
Self-adjointness is denoted as $X_i=X_i^\ast$ and $P_k=P_k^\ast$. 
\item[c)] The scalar product of a ket with itself belongs to 
$\mathbb{D}_+:=\{\alpha_{\wh{1}}\eb{1}{} 
+\alpha_{\wh{2}}\eb{2}{}: \alpha_{\wh{1}},
\alpha_{\wh{2}}\geq0\}$.
\item[d)] $\cro{X_i,P_k}=\ib{1}{}\hbar\del_{ik}\xi$, where $\xi\in\T$ is not in the null cone, $\hbar$ is Planck's reduced constant and $\del_{ik}$ is Kronecker's delta.
\item[e)] There are eigenkets $\ke{\psi_E}$ of $H$ which are not in the null cone and whose corresponding eigenvalues $E$ are not in the null cone.
\item[f)] Eigenkets $\ke{\psi_E}$ corresponding to a given eigenvalue $E$ span a 
finite-dimensional $\T$-module.
\item[g)] Two eigenkets $\ke{\psi_{E_i}},\ke{\psi_{E_j}}$ of $H$, not in $\mathcal{NC}$ and with $\p{E_i-E_j}$ not in $\mathcal{NC}$, are orthogonal.
\end{enumerate}

Assumption~(a) introduces the bicomplex generalization of the position, momentum and energy operators. 
With~(b) we impose, as in the standard case,
the self-adjointness of $X_i$ and $P_k$. The third general property of the scalar product, stated after
eq.~\eqref{fbih_eq:defprodscafctbi}, implies that $\p{\ke{\psi},\ke{\psi}}\in\mathbb{D}$. The more restrictive assumption~(c) is added so that a ket $\ke{\psi}\not\in\mathcal{NC}$ can always be normalized~\cite{lavoie2011finite}, through multiplication by $\p{\ke{\psi},\ke{\psi}}^{-1/2}$.

The second and third general properties of the scalar product imply that $\p{\al\ke{\psi},\ke{\phi}}=\al^\dagger\p{\ke{\psi},\ke{\phi}}$. This means that the eigenvalues of a
self-adjoint bicomplex operator (associated with an eigenket that is not in the null cone) are hyperbolic numbers \cite{rochon2006bicomplex}.

The simplest possible form of a bicomplex extension of the canonical commutation relations seems to be imbedded in~(d). This assumption entails that none of the operators $X_i$ and $P_k$ belongs to the null cone. Assumption~(e) implies that $H$ is not in $\mathcal{NC}$. Indeed if $H_{\bo{1}}$ vanished, for instance, $H$ could not have an eigenvalue with $E_{\wh{1}} \neq 0$. Making use of the
self-adjointness of $X_i$ and $P_k$ and the properties of the scalar product, one can show that the bicomplex number $\xi$ introduced in~(d) is in fact a hyperbolic number \cite{lavoie2010bicomplex}.

Assumption~(f) is not really necessary.  In effect it restricts the eigenvalues we will obtain to the discrete spectrum.  Assumption~(g) can be seen as contributing to the specification of the scalar product.  That assumption is not necessary to derive eigenvalues through the algebraic method, but it is needed to give structure to the $\T$-module of eigenkets.  In the differential equation method that will be used in section~\ref{eigenfctsofH}, (g) will in fact be derived.

We will now show that, without loss of generality, $\xi$ can be taken in $\mathbb{D}^+$.  Note that $\mathbb{D}^+$ differs from  $\mathbb{D}_+$ introduced in assumption~(c) in that in $\mathbb{D}^+$, vanishing values of $\alpha_{\wh{1}}$ and $\alpha_{\wh{2}}$ are excluded. To show that $\xi$ can be taken in $\mathbb{D}^+$, we show that a simple rescaling of $X_i$ and $P_k$ transforms the problem specified in assumptions~(a) to~(g) into an equivalent one, but with $\xi$ in~$\mathbb{D}^+$.

For $s=1,2$, let $\al_{\wh{s}}$ and $\be_{\wh{s}}$ be in $\C\de{\ib{1}{}}$ and nonzero.  Define $X_i^\prime$ and $P_k^\prime$ so that
\begin{align}
X_i:=\p{\al_{\wh{1}}\eb{1}{}+\al_{\wh{2}}\eb{2}{}}X_i^\prime \quad \mbox{and} \quad P_k:=\p{\be_{\wh{1}}\eb{1}{}+\be_{\wh{2}}\eb{2}{}}P_k^\prime.\label{fbih_eq:rescalingXiPk}
\end{align}
We have
\begin{align*}
\p{\ke{\phi},X_i\ke{\psi}}&=%\p{\ke{\psi},\acco{\al_{\wh{1}}\eb{1}{}+\al_{\wh{2}}\eb{2}{}}X_i^\prime\ke{\psi}}\\
\acco{\al_{\wh{1}}\eb{1}{}+\al_{\wh{2}}\eb{2}{}}\p{\ke{\phi},X_i^\prime\ke{\psi}},\\
\p{X_i\ke{\phi},\ke{\psi}}&=%\p{\acco{\al_{\wh{1}}\eb{1}{}+\al_{\wh{2}}\eb{2}{}}X_i^\prime\ke{\psi},\ke{\psi}}\\
\acco{\ol{\al_{\wh{1}}}\eb{1}{}+\ol{\al_{\wh{2}}}\eb{2}{}}\p{X_i^\prime\ke{\phi},\ke{\psi}}.
\end{align*}
Since $X_i$ is self-adjoint, the left-hand sides of these two expressions are equal.  The
right-hand sides must also be equal.  Using that and requiring $X_i^\prime$ to be
self-adjoint, we find
%
%% One column
%\begin{comment}
\begin{align*}
\acco{\al_{\wh{1}}\eb{1}{}+\al_{\wh{2}}\eb{2}{}}\p{\ke{\phi},X_i^\prime\ke{\psi}}
= \acco{\ol{\al_{\wh{1}}}\eb{1}{}+\ol{\al_{\wh{2}}}\eb{2}{}}\p{\ke{\phi},X_i^\prime\ke{\psi}} .
\end{align*}
%\end{comment}
%% Two columns
\begin{comment}
\begin{align*}
& \acco{\al_{\wh{1}}\eb{1}{}+\al_{\wh{2}}\eb{2}{}}\p{\ke{\phi},X_i^\prime\ke{\psi}} \\
& \qquad = \acco{\ol{\al_{\wh{1}}}\eb{1}{}+\ol{\al_{\wh{2}}}\eb{2}{}}\p{\ke{\phi},X_i^\prime\ke{\psi}} .
\end{align*}
\end{comment}
%
Since $X_i^\prime$ is not in~$\mathcal{NC}$, one can always find kets $\ke{\phi}$ and $\ke{\psi}$ such that $\p{\ke{\phi},X_i^\prime\ke{\psi}}$
is not in~$\mathcal{NC}$.  Therefore $\al_{\wh{s}}=\ol{\al_{\wh{s}}}$.  By a similar argument, $\be_{\wh{s}}=\ol{\be_{\wh{s}}}$. 

Let us now substitute \eqref{fbih_eq:rescalingXiPk} into~\eqref{fbih_eq:defHfromXiPk}. In the idempotent basis we have
\begin{align}
H = \sum_{s}\acco{\frac{1}{2\mu}\sum_k \be_{\widehat{s}}^2{P_{k\bo{s}}^\prime}^2
-Ze^2\cro{\sum_i \al_{\widehat{s}}^2{X_{i\bo{s}}^\prime}^2}^{-1/2}}\eb{s}{}.\label{fbih_eq:HInvariant1}
\end{align}
Sums over indices like $i$, $j$ and $k$ run from~1 to~3, whereas the
idempotent-basis index $s$ runs from~1 to~2.  We want $H$ to have a form similar to~\eqref{fbih_eq:defHfromXiPk} or, again in the idempotent basis,
\begin{align}
H = \sum_{s}\acco{\frac{1}{2\mu^\prime}\sum_k {P_{k\bo{s}}^\prime}^2
-\p{Ze^2}^\prime\cro{\sum_i {X_{i\bo{s}}^\prime}^2}^{-1/2}}\eb{s}{}.\label{fbih_eq:HInvariant2}
\end{align}
Comparing \eqref{fbih_eq:HInvariant1} and \eqref{fbih_eq:HInvariant2}, we see that $\mu^\prime=\mu/\be_{\wh{s}}^2$. This holds for both $s=1$ and $2$.  Hence $\be_{\wh{1}}^2=\be_{\wh{2}}^2$ or, equivalently, $\be_{\wh{1}}=\pm\be_{\wh{2}}$. Comparing 
again~\eqref{fbih_eq:HInvariant1} and~\eqref{fbih_eq:HInvariant2}, we have
\begin{align*}
\p{Ze^2}^\prime\cro{\sum_i {X_{i\bo{s}}^\prime}^2}^{-1/2}
&=Ze^2\cro{\sum_i \al_{\widehat{s}}^2{X_{i\bo{s}}^\prime}^2}^{-1/2}.
\end{align*}
Once more, the only way $\p{Ze^2}^\prime$ can be a real positive number is if $\al_{\wh{1}}^2=\al_{\wh{2}}^2$, or $\al_{\wh{1}}=\pm\al_{\wh{2}}$.

Now assumption~(d) and definition \eqref{fbih_eq:rescalingXiPk} allow us to write
%
%% One column
%\begin{comment}
\begin{align*}
\cro{X_i,P_k} =\ib{1}{}\hbar\del_{ik}\p{\xi_{\widehat{1}}\eb{1}{}+\xi_{\widehat{2}}\eb{2}{}} 
=\p{\al_{\widehat{1}}\be_{\widehat{1}}\eb{1}{}+\al_{\widehat{2}}\be_{\widehat{2}}\eb{2}{}}\cro{X_i^\prime,P_k^\prime} .
\end{align*}
%\end{comment}
%% Two columns
\begin{comment}
\begin{align*}
\cro{X_i,P_k} & =\ib{1}{}\hbar\del_{ik}\p{\xi_{\widehat{1}}\eb{1}{}+\xi_{\widehat{2}}\eb{2}{}} \\ 
& =\p{\al_{\widehat{1}}\be_{\widehat{1}}\eb{1}{}+\al_{\widehat{2}}\be_{\widehat{2}}\eb{2}{}}\cro{X_i^\prime,P_k^\prime} .
\end{align*}
\end{comment}
%
This implies that
%
%% One column
%\begin{comment}
\begin{align*}
\cro{X_i^\prime,P_k^\prime}&=\ib{1}{}\hbar\del_{ik}\p{\frac{\xi_{\widehat{1}}}{\al_{\widehat{1}}\be_{\widehat{1}}}\eb{1}{}+\frac{\xi_{\widehat{2}}}{\al_{\widehat{2}}\be_{\widehat{2}}}\eb{2}{}}
=:\ib{1}{}\hbar\del_{ik}\p{\xi_{\widehat{1}}^\prime\eb{1}{}+\xi_{\widehat{2}}^\prime\eb{2}{}}.
\end{align*}
%\end{comment}
%% Two columns
\begin{comment}
\begin{align*}
\cro{X_i^\prime,P_k^\prime}
&=\ib{1}{}\hbar\del_{ik}\p{\frac{\xi_{\widehat{1}}}{\al_{\widehat{1}}\be_{\widehat{1}}}\eb{1}{}+\frac{\xi_{\widehat{2}}}{\al_{\widehat{2}}\be_{\widehat{2}}}\eb{2}{}} \\
& =:\ib{1}{}\hbar\del_{ik}\p{\xi_{\widehat{1}}^\prime\eb{1}{}+\xi_{\widehat{2}}^\prime\eb{2}{}}.
\end{align*}
\end{comment}
%
Therefore, we can always choose $\al_{\wh{s}}$ and $\be_{\wh{s}}$ so that $\xi_{\wh{1}}^\prime$ and $\xi_{\wh{2}}^\prime$ are real and positive.  Moreover, we can rescale $\xi_{\wh{1}}^\prime$ to be~1, but we cannot in general rescale $\xi_{\wh{2}}^\prime$ so that $\xi_{\wh{2}}^\prime = \xi_{\wh{1}}^\prime$. We conclude that $H$ can always be written as in \eqref{fbih_eq:defHfromXiPk}, with the commutation relations between $X_i$ and $P_k$ given by
\begin{align}
\cro{X_i,P_k}=\ib{1}{}\hbar\del_{ik}\p{\xi_{\widehat{1}}\eb{1}{}+\xi_{\widehat{2}}\eb{2}{}},\qquad \xi_{\widehat{1}},\xi_{\widehat{2}}\in\R^+.\label{fbih_eq:commuXiPk}
\end{align}

%\newpage
\subsection{Eigenvalues of $H$}\label{eigenvalofH}
In this section, eigenvalues of $H$ are obtained through Pauli's algebraic method. This shows the unicity of the eigenvalues obtained, under 
assumptions (a)--(g).

Just as in standard quantum mechanics, we define the bicomplex angular momentum operator $\boldsymbol{L}:=\boldsymbol{R}\times\boldsymbol{P}$.
In terms of the Levi-Civita symbol we have
\begin{align}
L_i=\sum_{jk}\ep_{ijk}X_jP_k=\sum_s\acco{\sum_{jk}\ep_{ijk}X_{j\bo{s}}P_{k\bo{s}}}\eb{s}{}.
\end{align}
The bicomplex
Runge-Lenz vector is defined as
\begin{align*}
\boldsymbol{A}:=\frac{1}{2\mu}\p{\boldsymbol{P}\times\boldsymbol{L}-\boldsymbol{L}\times\boldsymbol{P}}-Ze^2\frac{\boldsymbol{R}}{R},
\end{align*}
which is equivalent to
%
%% One column
%\begin{comment}
\begin{align}
A_i=\sum_s\acco{\frac{1}{\mu}\p{\sum_{jk}\ep_{ijk}P_{j\bo{s}}L_{k\bo{s}}-\ib{1}{}\hbar\xi_{\wh{s}} P_{i\bo{s}}}-Ze^2\frac{X_{i\bo{s}}}{R_{\bo{s}}}}\eb{s}{}.
\end{align}
%\end{comment}
%% Two columns
\begin{comment}
\begin{align}
A_i &=\sum_s \scalebox{1.175}{\Bigg\{} \frac{1}{\mu} \left( 
\sum_{jk}\ep_{ijk}P_{j\bo{s}}L_{k\bo{s}}
-\ib{1}{}\hbar\xi_{\wh{s}} P_{i\bo{s}} \right) \notag \\
& \qquad\qquad -Ze^2\frac{X_{i\bo{s}}}{R_{\bo{s}}}
\scalebox{1.175}{\Bigg\}} \eb{s}{}.
\end{align}
\end{comment}
%

Let us write $\eta:=\hbar\xi$. We then observe that the commutator of $X_{i\bo{s}}$ and $P_{j\bo{s}}$, as well as the definitions of $L_{i\bo{s}}$ and $A_{j\bo{s}}$, are the same as the ones in the standard
quantum-mechanical case, except that $\hbar$ is everywhere replaced by $\eta_{\wh{s}}$. By an argument identical to the one in the standard
case~\cite{englefield1972group,greiner1994quantum}, we therefore obtain the following commutation relations and properties:
\begin{align}
\cro{A_{i\bo{s}},H_{\bo{s}}} &=0,\label{fbih_eq:14.99a}\\
\cro{L_{i\bo{s}},A_{j\bo{s}}} &=\ib{1}{}\eta_{\wh{s}}\sum_k\ep_{ijk}A_{k\bo{s}},\label{fbih_eq:14.99b}\\
\cro{A_{i\bo{s}},A_{j\bo{s}}} &=-\frac{2\ib{1}{}\eta_{\wh{s}}}{\mu}H_{\bo{s}}\sum_k\ep_{ijk}L_{k\bo{s}},\label{fbih_eq:14.99}\\
\sum_iL_{i\bo{s}}A_{i\bo{s}} &=0=\sum_iA_{i\bo{s}}L_{i\bo{s}},\label{fbih_eq:14.100a}\\
A_{\bo{s}}^2 &=\p{Ze^2}^2+\frac{2}{\mu}H_{\bo{s}}\acco{L_{\bo{s}}^2+ \eta_{\wh{s}}^2}.\label{fbih_eq:14.100}
\end{align}
Here $L_{\bo{s}}^2 = \sum_i L_{i\bo{s}}^2$ and $A_{\bo{s}}^2 = \sum_i A_{i\bo{s}}^2$. 
By properties of the idempotent basis, operators $H$, $L_i$, and $A_j$ satisfy similar relations
as~\eqref{fbih_eq:14.99a}--\eqref{fbih_eq:14.100}, with the index $\bo{s}$ deleted.  Note that, because
of~\eqref{fbih_eq:14.99}, operators $H_{\bo{s}}$, $L_{i\bo{s}}$, and $A_{j\bo{s}}$ do not make up a Lie algebra. They do, however, generate an
infinite-dimensional one.

To avoid having to work with (the bicomplex generalization of) an
infinite-dimensional algebra, we will restrict the action of $H$, $L_i$, and $A_j$ on the module~$\mathcal{F}_E$ corresponding to a given eigenvalue $E$ of $H$. By assumption~(f), $\mathcal{F}_E$ is
finite-dimensional.  Assuming as in~(e) that $E$ is not in $\mathcal{NC}$, we define three operators $\tilde{A}_i$ acting on $\mathcal{F}_E$ as
\begin{align}
\tilde{A}_i:=\sqrt{\frac{-\mu}{2E}}\,A_i.
\end{align}
Equations \eqref{fbih_eq:14.99a}--\eqref{fbih_eq:14.99} (without the index $\bo{s}$) imply quite straightforwardly that
\begin{align}
\cro{\tilde{A}_i,H} &=0,\label{fbih_eq:14.99_2a} \\
\cro{L_i,\tilde{A}_j} &=\ib{1}{}\eta\sum_k\ep_{ijk}\tilde{A}_k,\label{fbih_eq:14.99_2b} \\
\cro{\tilde{A}_i,\tilde{A}_j} &=\ib{1}{}\eta\sum_k\ep_{ijk}L_k. \label{fbih_eq:14.99_2}
\end{align}
Making use of \eqref{fbih_eq:14.100a} and~\eqref{fbih_eq:14.100}, we easily obtain
\begin{align}
\sum_iL_i\tilde{A}_i=0=\sum_i\tilde{A}_iL_i \label{fbih_eq:14.100p1}
\end{align}
and
\begin{align}
\sum_i\tilde{A}_i\tilde{A}_i+\sum_jL_jL_j+ \eta^2+\frac{\mu}{2E}\p{Ze^2}^2=0.\label{fbih_eq:14.100p2}
\end{align}

To arrive at explicit values of $E$, it is useful to construct six operators $F_i$ and $G_i$ as
\begin{align}
F_i:=\frac{1}{2}\p{L_i-\tilde{A}_i} \quad \mbox{and} \quad G_i:=\frac{1}{2}\p{L_i+\tilde{A}_i}.\label{fbih_eq:FietG_i}
\end{align}
Since $L_i$ and $\tilde{A}_i$ commute with $H$, so do $F_i$ and $G_i$. The commutation relations of the latter are given by
\begin{align}
\cro{F_i,F_j} &=\ib{1}{}\eta\sum_k\ep_{ijk}F_k,\label{fbih_eq:commuFietG_ja} \\
\cro{G_i,G_j} &=\ib{1}{}\eta\sum_k\ep_{ijk}G_k,\label{fbih_eq:commuFietG_jb} \\
\cro{F_i,G_j} &=0.\label{fbih_eq:commuFietG_j}
\end{align}
This means that the $F_i$ commute with the $G_j$, but the $F_i$ (and the $G_i$) have with themselves the same commutation relations as the bicomplex angular momentum operators. Since $F_i$ and $G_j$ commute, $F^2$ commutes with $G^2$. 
With~\eqref{fbih_eq:14.100p1} and~\eqref{fbih_eq:FietG_i}, one can show that $F^2-G^2=0$ and, therefore, that their eigenvalues are equal. 

If we project \eqref{fbih_eq:commuFietG_ja} on the idempotent basis, we find that for each $\bo{s}$ the $F_{i\bo{s}}$ have the commutation relations of standard angular momentum (with $\hbar$ replaced by $\eta_{\wh{s}}$). Eigenvalues of $F_{\bo{s}}^2$ are consequently equal to $f_{\wh{s}}\p{f_{\wh{s}}+1} \eta_{\wh{s}}^2$, where $f_{\wh{s}}$ is a nonnegative integer or half 
integer~\cite{edmonds1960angular}.  The eigenvalues of $F^2$ (and of $G^2$) are therefore equal to
\begin{align}
\sum_s f_{\wh{s}}\p{f_{\wh{s}}+1}\eta_{\wh{s}}^2 \, \eb{s}{} 
= f\p{f+1}\eta^2 ,
\end{align}
where of course $f = f_{\wh{1}} \eb{1}{} + f_{\wh{2}} \eb{2}{}$.

If we substitute \eqref{fbih_eq:FietG_i} in~\eqref{fbih_eq:14.100p2} we get
\begin{align*}
0=2\sum_iF_iF_i+2\sum_jG_jG_j+ \eta^2+\frac{\mu}{2E}\p{Ze^2}^2.
\end{align*}
For the eigenvalues this entails that 
\begin{align*}
0=\p{2f+1}^2 \eta^2+\frac{\mu}{2E}\p{Ze^2}^2
\end{align*}
or, in the idempotent basis,
\begin{align*}
0&=\sum_s\acco{\p{2f_{\wh{s}}+1}^2 \eta_{\wh{s}}^2+\frac{\mu}{2E_{\wh{s}}}\p{Ze^2}^2}\eb{s}{}.
\end{align*}
But then
\begin{align*}
\p{2f_{\wh{s}}+1}^2 \eta_{\wh{s}}^2+\frac{\mu}{2E_{\wh{s}}}\p{Ze^2}^2=0
\end{align*}
for both $s=1$ and 2. This yields for $E_{\wh{s}}$
%
%% One column
%\begin{comment}
\begin{align*}
E_{\wh{s}} 
=-\frac{\mu\p{Ze^2}^2}{2\eta_{\wh{s}}^2\p{2f_{\wh{s}}+1}^2} 
=-\frac{\mu\p{Ze^2}^2}{2\hbar^2\xi_{\wh{s}}^2\p{2f_{\wh{s}}+1}^2} =:-\frac{\mu Z^2e^4}{2\hbar^2\xi_{\wh{s}}^2n_{\wh{s}}^2},%\label{14.109}
\end{align*}
%\end{comment}
%% Two columns
\begin{comment}
\begin{align*}
E_{\wh{s}} 
&=-\frac{\mu\p{Ze^2}^2}{2\eta_{\wh{s}}^2\p{2f_{\wh{s}}+1}^2} 
=-\frac{\mu\p{Ze^2}^2}{2\hbar^2\xi_{\wh{s}}^2
\p{2f_{\wh{s}}+1}^2} \\ 
&=:-\frac{\mu Z^2e^4}{2\hbar^2\xi_{\wh{s}}^2n_{\wh{s}}^2},
%\label{14.109}
\end{align*}
\end{comment}
%
where $n_{\wh{s}}$ is a positive integer. This means that we can write
\begin{align}
E_n=-\frac{\mu Z^2e^4}{2\hbar^2\xi^2n^2}=\sum_s\acco{-\frac{\mu Z^2e^4}{2\hbar^2\xi_{\wh{s}}^2n_{\wh{s}}^2}}\eb{s}{}.\label{fbih_eq:EHBi}
\end{align}
This coincides with the standard Coulomb potential energy levels \cite{tannoudji1977quantum,marchildon2002quantum} if and only if $\xi_{\widehat{1}}=1=\xi_{\widehat{2}}$ and $n_{\widehat{1}}=n_{\widehat{2}}=n$. We call $n=n_{\wh{1}}\eb{1}{}+n_{\wh{2}}\eb{2}{}$ the bicomplex principal quantum number. In the hyperbolic representation,
%
%% One column
%\begin{comment}
\begin{align}
E_n=-\frac{\mu Z^2e^4}{4\hbar^2}\acco{\cro{\xi_{\wh{1}}^{-2}n_{\wh{1}}^{-2}+\xi_{\wh{2}}^{-2}n_{\wh{2}}^{-2}}+\cro{\xi_{\wh{1}}^{-2}n_{\wh{1}}^{-2}-\xi_{\wh{2}}^{-2}n_{\wh{2}}^{-2}}\jb{}{}}.\label{fbih_eq:EHhyper}
\end{align}
%\end{comment}
%% Two columns
\begin{comment}
\begin{align}
E_n &=-\frac{\mu Z^2e^4}{4\hbar^2}
\left\{ \cro{\xi_{\wh{1}}^{-2}n_{\wh{1}}^{-2}+\xi_{\wh{2}}^{-2}n_{\wh{2}}^{-2}} \right. \notag \\
& \qquad \left. +\cro{\xi_{\wh{1}}^{-2}n_{\wh{1}}^{-2}-\xi_{\wh{2}}^{-2}n_{\wh{2}}^{-2}}\jb{}{} \right\}.
\label{fbih_eq:EHhyper}
\end{align}
\end{comment}
%

We point out a formal symmetry satisfied 
by~\eqref{fbih_eq:EHhyper}.  Since $\jb{}{}:=\eb{1}{}-\eb{2}{}$, we can see that $\sqrt{\jb{}{}}=\eb{1}{}+\ib{1}{}\eb{2}{}$.  But then $\xi\sqrt{\jb{}{}} = \xi_{\wh{1}} \eb{1}{} + \ib{1}{} \xi_{\wh{2}} \eb{2}{}$. This immediately implies that
%
%% One column
%\begin{comment}
\begin{align*}
\Re{E_n,\xi}=\Hy{E_n, \xi\sqrt{\jb{}{}}} \quad \mbox{and} \quad \Re{E_n,\xi\sqrt{\jb{}{}}}=\Hy{E_n,\xi}.
\end{align*}
%\end{comment}
%% Two columns
\begin{comment}
\begin{align*}
\Re{E_n,\xi}&=\Hy{E_n, \xi\sqrt{\jb{}{}}} , \\
\Re{E_n,\xi\sqrt{\jb{}{}}}&=\Hy{E_n,\xi}.
\end{align*}
\end{comment}
%
Note, however, that $\xi\sqrt{\jb{}{}}$ is not a hyperbolic number.

%\newpage
\section{Eigenfunctions of $H$}\label{eigenfctsofH}

In this section, we define a 
coordinate-basis representation for the bicomplex operators $X_i$ and $P_k$. We then show that they, as well as $H$, are
self-adjoint with respect to the scalar product defined in section~\ref{normspace}. Finally, solving the hamiltonian eigenvalue equation in the coordinate basis, we recover eigenvalues given in~\eqref{fbih_eq:EHBi} and obtain the hyperbolic Coulomb potential eigenfunctions. 

\subsection{Coordinate-basis representation}\label{schrodeq}

We begin by constructing a representation of $X_i$ and $P_k$ on $\mathcal{F}_3$, the space of bicomplex
square-integrable functions on~$\R^3$. Letting $\bo{r}$ denote the triplet $(x_1,x_2,x_3)$, we define the action of $X_i$ as
\begin{align}
X_if\de{\bo{r}}:=x_if\de{\bo{r}}.\label{fbih_eq:repcoorX}
\end{align}
This implies that the $X_i$ commute two by two.  A function $F(\bo{R})$ acts on $f$ as
\begin{align}
F(\bo{R})f\de{\bo{r}}:=F(\bo{r})f\de{\bo{r}}.\label{fbih_eq:relfdeXifdexi}
\end{align}
For the action of $P_k$ we write, in a rather straightforward extension of the standard case
\begin{align}
P_kf\de{\bo{r}}:=-\ib{1}{}\hbar\xi\pD{}{x_k}f\de{\bo{r}}
=-\ib{1}{}\eta\pD{}{x_k}f\de{\bo{r}},\label{fbih_eq:repcoorP}
\end{align}
where $\xi = \xi_{\widehat{1}}\eb{1}{}+\xi_{\widehat{2}}\eb{2}{}$ and the $\xi_{\widehat{s}}$ are positive real 
numbers.  Strictly speaking, $P_k$ should be defined on a subset of $\mathcal{F}_3$, made up of suitably differentiable functions. We'll come back to this in section~\ref{veriandconclu}. Clearly the $P_k$ commute two by two.  Moreover, by letting both sides act on an arbitrary function $f$, one easily shows that
\begin{align}
&\cro{X_i,P_k} = \ib{1}{}\eta\del_{ik} .\label{momentum}
\end{align}

Let $f$ and $g$ be in $\mathcal{F}_3$.  Clearly $\p{X_i f,g}=\p{f,X_i g}$, so that $X_i$ is 
self-adjoint.  For $P_k$ we have 
\begin{align*}
&\p{P_k f,g}-\p{f,P_k g} \\
& \qquad = \ib{1}{}\eta^\dagger\p{\pD{f}{x_k},g}+\ib{1}{}\eta\p{f,\pD{g}{x_k}}\\
& \qquad =\ib{1}{}\eta\int\cro{\pD{f}{x_k}}^\dagger g\de{\bo{r}}\,\ud\boldsymbol{r}
+\ib{1}{}\eta\int f^{\dagger}
\de{\bo{r}}\pD{g}{x_k}\ud\boldsymbol{r}\\
& \qquad =\ib{1}{}\eta\int\pD{}{x_k}\cro{f^{\dagger}
\de{\bo{r}}g\de{\bo{r}}}\ud\boldsymbol{r} =0.
\end{align*}
To obtain the last equality, we have restricted the space of functions to those that vanish at infinity and on these functions, $P_k$ is 
self-adjoint.  The proof that $\p{Hf,g}=\p{f,Hg}$, and therefore that $H$ is
self-adjoint, is straightforward.

%\newpage
\subsection{Wave functions}\label{eigenfctofH}

The bicomplex quantum Coulomb potential problem consists in solving the 
three-dimensional eigenvalue equation
\begin{align}
H\psi_E\de{\bo{r}}=E\psi_E\de{\bo{r}}\label{fbih_eq:eqSpaValPropre}
\end{align}
for $H$ given by~\eqref{fbih_eq:defHfromXiPk}. Making use
of~\eqref{fbih_eq:relfdeXifdexi}
and~\eqref{fbih_eq:repcoorP} we can write more explicitly
\begin{align}
- \left\{\frac{\eta^2}{2\mu} \nabla^2 + \frac{Ze^2}{r} \right\} \psi_E\de{\bo{r}}
= E\psi_E\de{\bo{r}}\label{fbih_eq:eqSpaValPropre1} .
\end{align}

We now write $\xi$, $E$, and $\psi_E$ in the idempotent basis.  Equation~\eqref{fbih_eq:eqSpaValPropre1} becomes
\begin{align}
\sum_s \left\{\frac{\eta_{\wh{s}}^2}{2\mu} \nabla^2 + \frac{Ze^2}{r} + E_{\wh{s}} \right\}
(\psi_E)_{\wh{s}}\de{\bo{r}} \eb{s}{} = 0 \label{fbih_eq:eqSpaValPropre2} .
\end{align}
Clearly, each coefficient of $\eb{s}{}$ must separately vanish.  Writing the laplacian in spherical coordinates and making use of the standard expression of the angular momentum operator we get
\begin{align}
\acco{\frac{\eta^2_{\wh{s}}}{2\mu}\cro{\frac{1}{r^2}\pD{}{r}\p{r^2\pD{}{r}}-\frac{1}{r^2}\frac{L_{\wh{s}}^2}{\eta^2_{\wh{s}}}}+\frac{Ze^2}{r}+E_{\wh{s}}} (\psi_E)_{\wh{s}}=0,\label{fbih_eq:eqValPropre3D}
\end{align}
for $s=1$ and $2$. Now we know from standard quantum mechanics that the spherical harmonics are eigenfunctions of the square of the angular momentum, that is, 
\begin{align}
L_{\wh{s}}^2Y_{l_{\wh{s}}m_{\wh{s}}}=l_{\wh{s}}\p{l_{\wh{s}}+1}\eta_{\wh{s}}^2Y_{l_{\wh{s}}m_{\wh{s}}},\label{fbih_eq:ActionL2harmo}
\end{align}
with $-l_{\wh{s}}\leq m_{\wh{s}}\leq l_{\wh{s}}$. Looking for solutions of the form
\begin{align}
(\psi_E)_{\wh{s}}\de{r,\theta,\phi}:=u_{\wh{s}}\de{r}Y_{l_{\wh{s}}m_{\wh{s}}}\de{\theta,\phi}
\end{align}
and using \eqref{fbih_eq:ActionL2harmo}, we get for $s=1, 2$
%
%% One column
%\begin{comment}
\begin{align}
\frac{1}{r^2}\D{}{r}\cro{r^2\D{}{r}u_{\wh{s}}\de{r}}-\cro{\frac{l_{\wh{s}}\p{l_{\wh{s}}+1}}{r^2}-\frac{2\mu}{\eta_{\wh{s}}^2}\p{\frac{Ze^2}{r}+E_{\wh{s}}}}u_{\wh{s}}\de{r}=0.\label{fbih_eq:eqrad}
\end{align}
%\end{comment}
%% Two columns
\begin{comment}
\begin{align}
& \frac{1}{r^2}\D{}{r}\cro{r^2\D{}{r}u_{\wh{s}}\de{r}} \notag \\
& \qquad -\cro{\frac{l_{\wh{s}}\p{l_{\wh{s}}+1}}{r^2}-\frac{2\mu}{\eta_{\wh{s}}^2}\p{\frac{Ze^2}{r}+E_{\wh{s}}}}u_{\wh{s}}\de{r}=0.\label{fbih_eq:eqrad}
\end{align}
\end{comment}
%
Just as in the standard case, the suitably normalized solutions of~\eqref{fbih_eq:eqrad} are given 
by~\cite{tannoudji1977quantum,marchildon2002quantum}
%
%% One column
%\begin{comment}
\begin{align}
u_{n_{\wh{s}}l_{\wh{s}}}\de{r}=\cro{\p{\frac{2Z}{n_{\wh{s}}a_{\wh{s}0}}}^3\frac{\p{n_{\wh{s}}-l_{\wh{s}}-1}!}{2n_{\wh{s}}\cro{\p{n_{\wh{s}}+l_{\wh{s}}}!}}}^{1/2}\ue^{-\ze_{\wh{s}}/2}\ze_{\wh{s}}^{l_{\wh{s}}}L_{n_{\wh{s}}-l_{\wh{s}}-1}^{2l_{\wh{s}}+1}\de{\ze_{\wh{s}}},\label{fbih_eq:unklk}
\end{align}
%\end{comment}
%% Two columns
\begin{comment}
\begin{align}
u_{n_{\wh{s}}l_{\wh{s}}}\de{r}
&=\cro{\p{\frac{2Z}{n_{\wh{s}}a_{\wh{s}0}}}^3\frac{\p{n_{\wh{s}}-l_{\wh{s}}-1}!}{2n_{\wh{s}}\cro{\p{n_{\wh{s}}+l_{\wh{s}}}!}}}^{1/2} \notag \\
& \qquad\qquad \cdot \ue^{-\ze_{\wh{s}}/2}\ze_{\wh{s}}^{l_{\wh{s}}}L_{n_{\wh{s}}-l_{\wh{s}}-1}^{2l_{\wh{s}}+1}
\de{\ze_{\wh{s}}},\label{fbih_eq:unklk}
\end{align}
\end{comment}
%
where $l_{\wh{s}} < n_{\wh{s}}$, the $L_{n_{\wh{s}}-l_{\wh{s}}-1}^{2l_{\wh{s}}+1}$ are Laguerre polynomials and
\begin{align}
\ze_{\wh{s}}:=\frac{2Z}{n_{\wh{s}}a_{0\wh{s}}}r, \qquad a_{0\wh{s}}:=a_0\xi_{\wh{s}}^2=\frac{\eta_{\wh{s}}^2}{\mu e^2}.\label{fbih_eq:7point20}
\end{align}
Each solution corresponds to an $E_{\wh{s}}$ given by $-\mu Z^2e^4/2\eta_{\wh{s}}^2n_{\wh{s}}^2$. Thus we recover the
eigenvalues~\eqref{fbih_eq:EHBi}, whose degeneracy is equal to the product $n_{\wh{1}}^2 n_{\wh{2}}^2$ of standard Coulomb potential degeneracies. The bicomplex wave functions
in~\eqref{fbih_eq:eqSpaValPropre} can be written as
\begin{align}
\psi_{nlm}\de{\bo{r}} = u_{nl} (r) Y_{lm} (\theta,\phi)
= \sum_s u_{n_{\wh{s}}l_{\wh{s}}}\de{r}Y_{l_{\wh{s}}m_{\wh{s}}}\de{\theta,\phi}\eb{s}{}.\label{fbih_eq:psinlmbi}
\end{align}

For $\xi_{\wh{1}}$ and $\xi_{\wh{2}}$ fixed,
any sextuplet $\p{n_{\wh{1}},n_{\wh{2}},l_{\wh{1}},l_{\wh{2}},m_{\wh{1}},m_{\wh{2}}}$ defines an eigenfunction of $H$. All functions with the same ($n_{\wh{1}}, n_{\wh{2}}$) correspond to the same eigenvalue.  A general eigenfunction of $H$ can therefore be written as
\begin{align}
\sum_{s}\sum_{l_{\wh{s}}=0}^{n_{\wh{s}}-1}\sum_{m_{\wh{s}}=-l_{\wh{s}}}^{l_{\wh{s}}}C_{l_{\wh{s}}m_{\wh{s}}}u_{n_{\wh{s}}l_{\wh{s}}}\de{r}Y_{l_{\wh{s}}m_{\wh{s}}}\de{\theta,\phi}\eb{s}{} \label{fbih_eq:solgenpsinlmbi} ,
\end{align}
with $C_{l_{\wh{s}}m_{\wh{s}}}\in\C\de{\ib{1}{}}$.
Along the way, we have introduced the bicomplex orbital quantum number $l=l_{\wh{1}}\eb{1}{}+l_{\wh{2}}\eb{2}{}$ and the bicomplex magnetic quantum number $m=m_{\wh{1}}\eb{1}{}+m_{\wh{2}}\eb{2}{}$. When $n_{\wh{1}} = n_{\wh{2}}$, the number $n$ is real, and similarly with $l$ and~$m$.

%\newpage
\section{Graphical representation of related functions}\label{graphrep}

Let us now go back to the eigenfunctions~\eqref{fbih_eq:psinlmbi} and consider their radial part only.  We write
\begin{align}
u_{nl}\de{r}:=u_{n_{\wh{1}}l_{\wh{1}}}\eb{1}{}+u_{n_{\wh{2}}l_{\wh{2}}}\eb{2}{}.\label{fbih_eq:fctradbi}
\end{align}
It is instructive to use the decomposition of $\acco{\eb{1}{},\eb{2}{}}$ in terms of $\jb{}{}$ given in
section~\ref{algstruc}.  We can then 
rewrite~\eqref{fbih_eq:fctradbi} as
%
%% One column
%\begin{comment}
\begin{align}
u_{nl}\de{r} & =\frac{1}{2} \sum_s \sqrt{u_{n_{\wh{s}}l_{\wh{s}}}^0}\,\xi_{\wh{s}}^{-3}\ue^{-\ze_{\wh{s}}/2}\ze_{\wh{s}}^{l_{\wh{s}}}L_{n_{\wh{s}}-l_{\wh{s}}-1}^{2l_{\wh{s}}+1}\de{\ze_{\wh{s}}} \notag \\
& \qquad + \frac{1}{2} \sum_s \p{-1}^{s+1}\sqrt{u_{n_{\wh{s}}l_{\wh{s}}}^0}\,\xi_{\wh{s}}^{-3}\ue^{-\ze_{\wh{s}}/2}\ze_{\wh{s}}^{l_{\wh{s}}}L_{n_{\wh{s}}-l_{\wh{s}}-1}^{2l_{\wh{s}}+1}\de{\ze_{\wh{s}}}\jb{}{} ,\label{fbih_eq:uhyrep}
\end{align}
%\end{comment}
%% Two columns
\begin{comment}
\begin{align}
& u_{nl}\de{r} =\frac{1}{2} \sum_s \sqrt{u_{n_{\wh{s}}l_{\wh{s}}}^0}\,\xi_{\wh{s}}^{-3}\ue^{-\ze_{\wh{s}}/2}\ze_{\wh{s}}^{l_{\wh{s}}}L_{n_{\wh{s}}-l_{\wh{s}}-1}^{2l_{\wh{s}}+1}\de{\ze_{\wh{s}}} \notag \\
& \quad + \frac{1}{2} \sum_s \p{-1}^{s+1}\sqrt{u_{n_{\wh{s}}l_{\wh{s}}}^0}\,\xi_{\wh{s}}^{-3}\ue^{-\ze_{\wh{s}}/2}\ze_{\wh{s}}^{l_{\wh{s}}}L_{n_{\wh{s}}-l_{\wh{s}}-1}^{2l_{\wh{s}}+1}\de{\ze_{\wh{s}}}\jb{}{} ,\label{fbih_eq:uhyrep}
\end{align}
\end{comment}
%
where $\sqrt{u_{n_{\wh{s}}l_{\wh{s}}}^0}\,\xi_{\wh{s}}^{-3}$ is the normalization constant.

It will be useful to define, as 
in~\eqref{eqalpha},
\begin{align}
\ze_{\wh{1}}=x_{\ze}+y_{\ze} , \qquad
\ze_{\wh{2}}=x_{\ze}-y_{\ze} \label{varzeta} ,
\end{align}
and consider the three functions $\Re{u_{nl}}$, $\Hy{u_{nl}}$ and $\abs{u_{nl}}$ as depending on $r$ through the two variables $x_{\ze}$ and $y_{\ze}$. With  $\xi_{\wh{s}}$ fixed, graphical representations of these functions can easily be obtained by assigning specific values to $n_{\wh{s}}$ and $l_{\wh{s}}$.

\begin{figure}[h!]
\centering
\subfloat[$a_0^{3/2}\Re{u_{25251212}}$.]{\label{gr2a}\includegraphics[width=0.35\columnwidth]{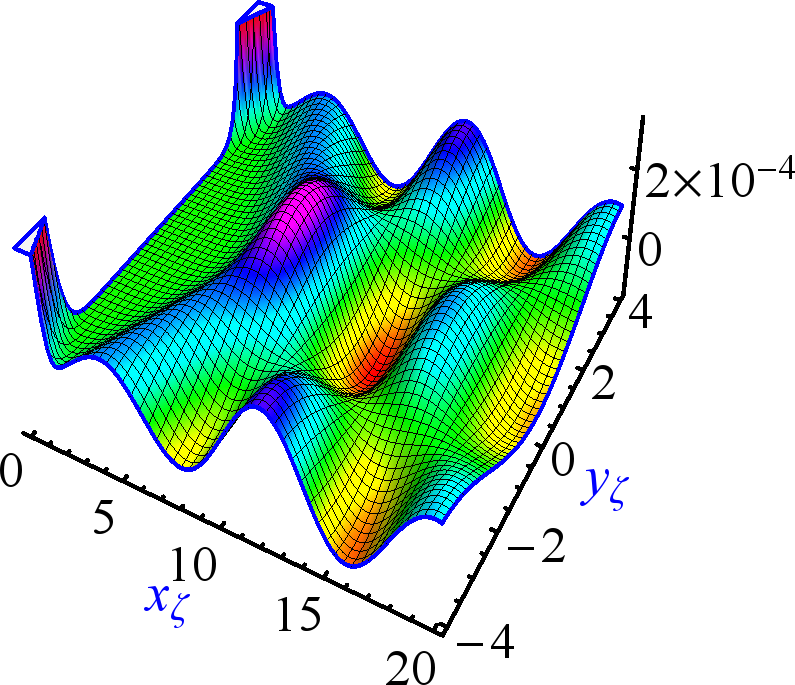}}\\%*[5ex]
\subfloat[$a_0^{3/2}\Hy{u_{25251212}}$.]{\label{gr2b}\includegraphics[width=0.35\columnwidth]{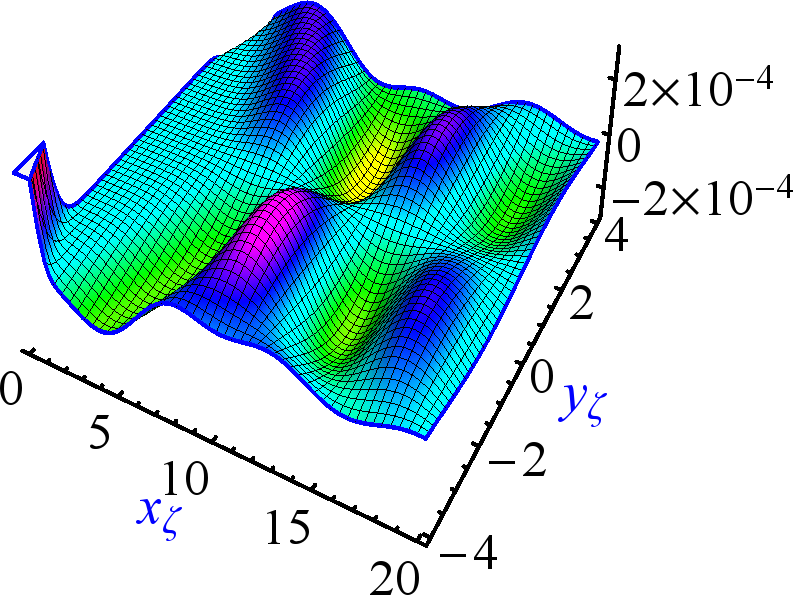}}\\%*[5ex]
\subfloat[$a_0^{3}\abs{u_{25251212}}^2$.]{\label{gr2c}\includegraphics[width=0.35\columnwidth]{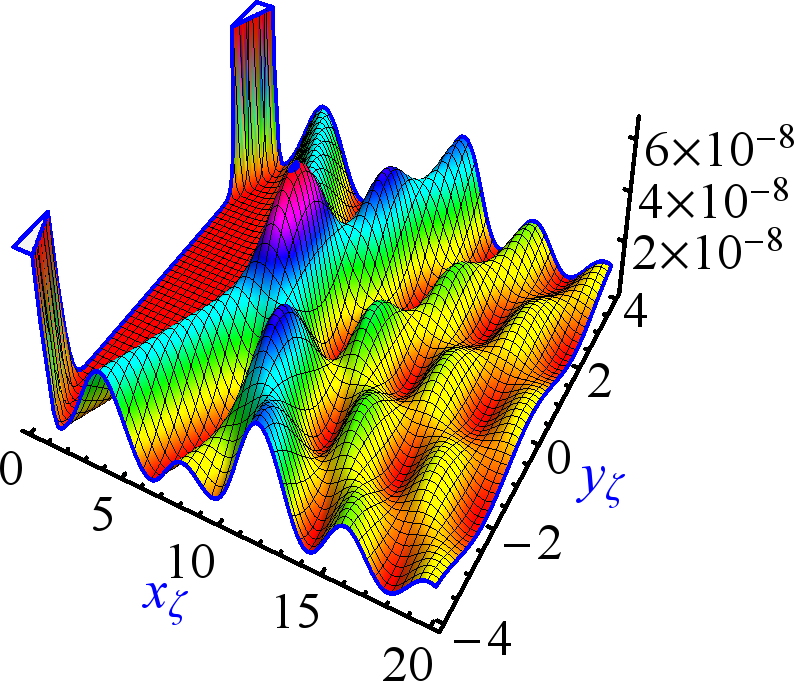}}
{\caption[]{$u_{nl}$ as a function of two independent variables.}\label{gr2}}
\end{figure}

In figure~\ref{gr2}, however, we go beyond the representation of eigenfunctions of $H$ and extend $x_{\zeta}$ and $y_{\zeta}$ to genuinely independent variables.  For $\xi$ fixed, this is equivalent to considering $r$ in the hyperbolic plane.  This allows for interesting surfaces to emerge, connected to a new class of nontrivial polynomials in two real variables.  For illustration, we take $\xi_{\wh{1}} = 1 = \xi_{\wh{2}}$ and let $n_{\wh{s}} = 25$ and $l_{\wh{s}} = 12$ for $s=1, 2$. In these plots, a cut at $y_{\ze}=0$ makes the hyperbolic part of $u_{nl}$ vanish.  The real part and real norm then coincide with the ones in the standard case. 

There is another equivalent way to generate the radial surfaces in the hyperbolic basis $\acco{1,\jb{}{}}$ in the case where $\xi_{\wh{1}} = 1 = \xi_{\wh{2}}$. We can write 
eigenfunctions~\eqref{fbih_eq:fctradbi} as
\begin{align*}
u_{nl}=\sqrt{u_{nl}^0}\,\xi^{-3}\ue^{-\ze/2}\ell_{nl}\de{\ze},
\end{align*}
where $\xi:=x_{\xi}+y_{\xi}\jb{}{}$,
$\ze:=x_{\ze}+y_{\ze}\jb{}{}$ and
%
%% One column
%\begin{comment}
\begin{align*}
\ell_{nl}\de{\ze}:=\ze^{l} L_{n-l-1}^{2l+1}\de{\ze} =\Re{\ell_{nl}\de{x_{\ze},y_{\ze}}}+\Hy{\ell_{nl}\de{x_{\ze},y_{\ze}}}\jb{}{},
\end{align*}
%\end{comment}
%% Two columns
\begin{comment}
\begin{align*}
\ell_{nl}\de{\ze}
&:=\ze^{l} L_{n-l-1}^{2l+1}\de{\ze} \\
&=\Re{\ell_{nl}\de{x_{\ze},y_{\ze}}}
+\Hy{\ell_{nl}\de{x_{\ze},y_{\ze}}}\jb{}{},
\end{align*}
\end{comment}
%
where
\begin{align*}
L_{n-l-1}^{2l+1}\de{\ze} := \sum_s 
L_{n_{\wh{s}}-l_{\wh{s}}-1}^{2l_{\wh{s}}+1}\de{\ze_{\wh{s}}} 
\eb{s}{} .
\end{align*}
Using the correspondence between the idempotent and hyperbolic bases, one can easily show that
%
%% One column
%\begin{comment}
\begin{align*}
\xi^{-3}&=\frac{1}{2}\acco{\p{x_{\xi}+y_{\xi}}^{-3}+\p{x_{\xi}-y_{\xi}}^{-3}}+\frac{1}{2}\acco{\p{x_{\xi}+y_{\xi}}^{-3}-\p{x_{\xi}-y_{\xi}}^{-3}}\jb{}{},
\end{align*}
%\end{comment}
%% Two columns
\begin{comment}
\begin{align*}
\xi^{-3}&=\frac{1}{2}\acco{\p{x_{\xi}+y_{\xi}}^{-3}+\p{x_{\xi}-y_{\xi}}^{-3}} \\
& \qquad +\frac{1}{2}\acco{\p{x_{\xi}+y_{\xi}}^{-3}-\p{x_{\xi}-y_{\xi}}^{-3}}\jb{}{},
\end{align*}
\end{comment}
%
which we define as $x_{\xi}^\prime+y_{\xi}^\prime\jb{}{}$. It is not difficult to show that the exponential transforms as
\begin{align*}
\ue^{-\ze/2}%&=\ue^{-x_{\ze}/2}\expde{-\frac{y_{\ze}}{2}\jb{}{}}=\ue^{-x_{\ze}/2}\acco{\cosh\p{-\frac{y_{\ze}}{2}}+\jb{}{}\sinh\p{-\frac{y_{\ze}}{2}}}\\
=\ue^{-x_{\ze}/2}\acco{\cosh\frac{y_{\ze}}{2}-\jb{}{}\sinh\frac{y_{\ze}}{2}}.
\end{align*}
This suggests that we can explicitly write
%
%% One column
%\begin{comment} 
\begin{align*}
\Re{u_{nl}}&=\sqrt{u_{nl}^0}\ue^{-x_{\ze}/2}\bigg[\p{x_{\xi}^\prime\cosh\frac{y_{\ze}}{2}
-y_{\xi}^\prime\sinh\frac{y_{\ze}}{2}}\Re{\ell_{nl}\de{x_{\ze},y_{\ze}}}\\
& \qquad +\p{y_{\xi}^\prime\cosh\frac{y_{\ze}}{2}-x_{\xi}^\prime\sinh\frac{y_{\ze}}{2}}\Hy{\ell_{nl}\de{x_{\ze},y_{\ze}}}\bigg],\\
\Hy{u_{nl}}&=\sqrt{u_{nl}^0}\ue^{-x_{\ze}/2}\bigg[\p{x_{\xi}^\prime\cosh\frac{y_{\ze}}{2}
-y_{\xi}^\prime\sinh\frac{y_{\ze}}{2}}\Hy{\ell_{nl}\de{x_{\ze},y_{\ze}}}\\
& \qquad +\p{y_{\xi}^\prime\cosh\frac{y_{\ze}}{2}-x_{\xi}^\prime\sinh\frac{y_{\ze}}{2}}\Re{\ell_{nl}\de{x_{\ze},y_{\ze}}}\bigg].
\end{align*}
%\end{comment}
%% Two columns
\begin{comment} 
\begin{align*}
&\Re{u_{nl}}=\sqrt{u_{nl}^0}\ue^{-x_{\ze}/2} \\
& \qquad \cdot \bigg[\p{x_{\xi}^\prime\cosh\frac{y_{\ze}}{2}
-y_{\xi}^\prime\sinh\frac{y_{\ze}}{2}}\Re{\ell_{nl}\de{x_{\ze},y_{\ze}}}\\
& \qquad\qquad +\p{y_{\xi}^\prime\cosh\frac{y_{\ze}}{2}-x_{\xi}^\prime\sinh\frac{y_{\ze}}{2}}\Hy{\ell_{nl}\de{x_{\ze},y_{\ze}}}\bigg]
\end{align*}
and
\begin{align*}
&\Hy{u_{nl}}=\sqrt{u_{nl}^0}\ue^{-x_{\ze}/2}\\
& \qquad \cdot \bigg[\p{x_{\xi}^\prime\cosh\frac{y_{\ze}}{2}
-y_{\xi}^\prime\sinh\frac{y_{\ze}}{2}}\Hy{\ell_{nl}\de{x_{\ze},y_{\ze}}}\\
& \qquad\qquad +\p{y_{\xi}^\prime\cosh\frac{y_{\ze}}{2}-x_{\xi}^\prime\sinh\frac{y_{\ze}}{2}}\Re{\ell_{nl}\de{x_{\ze},y_{\ze}}}\bigg].
\end{align*}
\end{comment}
%
With a symbolic computation software, we can generate the polynomial $\ell_{nl}\de{\ze}$ for any positive integers 
$n_{\wh{s}}$ and $l_{\wh{s}} \leq n_{\wh{s}}-1$, for $s=1, 2$. We then take $n=25$, $l=12$, $x_{\xi}=1$, $y_{\xi}=0$ (implying that $x_{\xi}^\prime=1$ and $y_{\xi}^\prime=0$), compute $\ell_{nl}\de{\ze}$, transform $\ze$ into $x_{\ze}+y_{\ze}\jb{}{}$ and separate the real and hyperbolic parts to get $\Re{\ell_{nl}\de{x_{\ze},y_{\ze}}}$ and $\Hy{\ell_{nl}\de{x_{\ze},y_{\ze}}}$ explicitly. If we plot the associated $a_0^{3/2}\Re{u}$, $a_0^{3/2}\Hy{u}$ and $a_0^{3}\abs{u}^2$, we recover the results shown in figure~\ref{gr2}.

%\newpage
\section{Discussion}\label{veriandconclu}

We have solved the eigenvalue equation~\eqref{fbih_eq:eqSpaValPropre} for the discrete spectrum of the Coulomb potential
hamiltonian~\eqref{fbih_eq:defHfromXiPk} in the framework of bicomplex numbers. The continuous spectrum could also be worked out along similar lines. The eigenvalues corresponding to the discrete spectrum are given
in~\eqref{fbih_eq:EHBi} and the eigenfunctions
in~\eqref{fbih_eq:psinlmbi}. Note that if
$\xi_{\wh{1}} = 1 = \xi_{\wh{2}}$, the standard wave functions  can be recovered by letting $n_{\wh{1}}=n_{\wh{2}}$, $l_{\wh{1}}=l_{\wh{2}}$ and $m_{\wh{1}}=m_{\wh{2}}$.

It is instructive to investigate the orthogonality properties of the
eigenfunctions~\eqref{fbih_eq:psinlmbi}.  Making use of
definition~\eqref{fbih_eq:defprodscafctbi} of the scalar product, one can write
\begin{align}
\p{\psi_{nlm},\psi_{n^\prime l^\prime m^\prime}}
&=\int\psi_{nlm}^\dagger\de{\bo{r}}\psi_{n^\prime l^\prime m^\prime}\de{\bo{r}}\ud\bo{r} \notag \\
&=\sum_s\eb{s}{} \int\psi_{n_{\wh{s}} l_{\wh{s}} m_{\wh{s}}}^\dagger\de{\bo{r}} 
\psi_{n_{\wh{s}}^\prime l_{\wh{s}}^\prime m_{\wh{s}}^\prime}\de{\bo{r}}\ud\bo{r} .\label{ortho}
\end{align}
It is well-known~\cite{tannoudji1977quantum,marchildon2002quantum} that the standard Coulomb problem eigenfunctions are orthonormal in all indices.  This implies that
\begin{align}
\p{\psi_{nlm},\psi_{n^\prime l^\prime m^\prime}}
=\sum_s\eb{s}{} \delta_{n_{\wh{s}} n_{\wh{s}}^\prime} \delta_{l_{\wh{s}} l_{\wh{s}}^\prime} \delta_{m_{\wh{s}} m_{\wh{s}}^\prime} .\label{ortho1}
\end{align}
From~\eqref{ortho1} we can draw two conclusions:
\begin{enumerate}
\item The eigenfunction $\psi_{nlm}$ is normalized.  Indeed
\begin{align}
\p{\psi_{nlm},\psi_{nlm}}
=\sum_s\eb{s}{} \delta_{n_{\wh{s}} n_{\wh{s}}} \delta_{l_{\wh{s}} l_{\wh{s}}} \delta_{m_{\wh{s}} m_{\wh{s}}}
=\sum_s\eb{s}{} = 1 .
\end{align}
\item If $E_n - E_{n^\prime}$ is not in the null cone, then $\psi_{nlm}$ and $\psi_{n^\prime l^\prime m^\prime}$ are orthogonal.  Indeed
from~\eqref{fbih_eq:EHBi} we see that $E_n - E_{n^\prime}$ is not in the null cone if and only if $n_{\wh{s}} \neq n_{\wh{s}}^\prime$ for $s=1, 2$.  But then $\delta_{n_{\wh{s}} n_{\wh{s}}^\prime} = 0$ for $s=1, 2$ and the orthogonality follows
from~\eqref{ortho1}.
\end{enumerate}

Let us consider the set of all finite linear combinations of eigenfunctions $\psi_{nlm}$, with bicomplex coefficients $C_{nlm}$.  It is straightforward to show that this set makes up a $\T$-module, which we denote as~$\mathcal{M}$.  Defining $X_i$ and $P_k$ as
in~\eqref{fbih_eq:repcoorX} and~\eqref{fbih_eq:repcoorP}, one sees that these operators are well-defined on~$\mathcal{M}$.  Moreover, it is not difficult to show that properties (a)--(g) in section~\ref{defsandaxioms} are satisfied in~$\mathcal{M}$, thus proving their consistency.

Of course, $\mathcal{M}$ is not a Hilbert space, since the restriction to finite linear combinations entails that it is not complete.  It is well-known that the Coulomb potential eigenfunctions in standard quantum mechanics, \ie the $\psi_{n_{\wh{s}} l_{\wh{s}} m_{\wh{s}}}$, make up an orthonormal system in the Hilbert space $L^2(\R^3)$~\cite{prugovecki}.
From~\cite{lavoie2010infinite}, one concludes that the $\psi_{nlm}$ make up an orthonormal system in a bicomplex Hilbert 
space $L^2(\R^3) \oplus L^2(\R^3)$.

We close this section with a result that we prove with the notation of the Coulomb problem, but that clearly holds more generally.  Let $U$ be a bicomplex linear operator, acting on~$\mathcal{M}$, that commutes with $H$. Then
\begin{align}
\p{E_n-E_{n^\prime}} \p{\psi_{nlm},U \psi_{n^\prime l^\prime m^\prime}}=0. \label{result6}
\end{align}
The proof is straightforward:
\begin{align*}
0 &=\p{\psi_{nlm},\cro{H,U}\psi_{n^\prime l^\prime m^\prime}}\\
&=\p{\psi_{nlm},\acco{HU-UH}\psi_{n^\prime l^\prime m^\prime}}\\
&=\p{\psi_{nlm},HU\psi_{n^\prime l^\prime m^\prime}}-\p{\psi_{nlm},UH\psi_{n^\prime l^\prime m^\prime}}\\
&=\p{U^\ast H^\ast\psi_{nlm},\psi_{n^\prime l^\prime m^\prime}}-\p{\psi_{nlm},UH\psi_{n^\prime l^\prime m^\prime}}\\
&=E_n\p{U^\ast\psi_{nlm},\psi_{n^\prime l^\prime m^\prime}}-E_{n^\prime}\p{\psi_{nlm},U\psi_{n^\prime l^\prime m^\prime}}\\
&=\p{E_n-E_{n^\prime}}\p{\psi_{nlm},U\psi_{n^\prime l^\prime m^\prime}}.
\end{align*}
This means that if $E_n-E_{n^\prime}$ is not in the null cone, then $\p{\psi_{nlm},U \psi_{n^\prime l^\prime m^\prime}}$ vanishes.  In other words,
$U \psi_{n^\prime l^\prime m^\prime}$ is a linear combination of functions associated with eigenvalue $E_{n^\prime}$.

%\newpage
\section{Conclusion}\label{conclu}

We have shown that, just like the quantum harmonic oscillator problem~\cite{lavoie2010bicomplex}, the quantum Coulomb problem can be solved in the framework of bicomplex numbers.  We have obtained the eigenvalues and eigenfunctions of the bicomplex hamiltonian, and have shown that the eigenfunctions make up an orthonormal system in a bicomplex Hilbert space.  The question is still open whether the constants $\xi_{\wh{1}}$ and $\xi_{\wh{2}}$ can be given a physical interpretation.  In any case, it is likely that the mathematical properties of the functions we introduced can be fruitfully studied for their own sake.

\section*{Acknowledgments}\label{acknowledg}

LM and DR are grateful to the Natural Sciences and Engineering Research Council of Canada for financial support. 

%\newpage

\end{document}